\documentclass[twocolumn,trackchanges]{aastex63}

\newcommand{\msun}{\rm M_{\odot}}

\shorttitle{MADs in Quiescent BHB: Formation, Signatures, and PeVatrons}
\shortauthors{Kimura, Sudoh, Kashiyama, Kawanaka}

\begin{document}
\title{Magnetically Arrested Disks in Quiescent Black-Hole Binaries: \\
Formation Scenario, Observable Signatures, and Potential PeVatrons}

\correspondingauthor{Shigeo S. Kimura}
\email{shigeo@astr.tohoku.ac.jp}

\author[0000-0003-2579-7266]{Shigeo S. Kimura}
\altaffiliation{JSPS Fellow}
\affiliation{Frontier Research Institute for Interdisciplinary Sciences, Tohoku University, Sendai 980-8578, Japan}
\affiliation{Astronomical Institute, Graduate School of Science, Tohoku University, Sendai 980-8578, Japan}

\author[0000-0002-6884-1733]{Takahiro Sudoh}
\affiliation{Department of Astronomy, University of Tokyo, Hongo, Tokyo 113-0033, Japan}
\author[0000-0003-4299-8799]{Kazumi Kashiyama}
\affil{Research Center for the Early Universe, Graduate School of Science, University of Tokyo, Bunkyo-ku, Tokyo 113-0033, Japan}
\affil{Department of Physics, Graduate School of Science, University of Tokyo, Bunkyo-ku, Tokyo 113-0033, Japan}
\author[0000-0001-8181-7511]{Norita Kawanaka}
\affiliation{Department of Astronomy, Graduate School of Science, Kyoto University, Kitashirakawa Oiwake-cho, Sakyo-ku, Kyoto, 606-8502, Japan}
\affiliation{Hakubi Center, Kyoto University, Yoshida-honmachi, Sakyo-ku, Kyoto, 606-8501, Japan}




\begin{abstract}
We propose magnetically arrested disks (MADs) in quiescent (low luminosity) black-hole (BH) binaries as the origin of the multiwavelength emission, and argue that this class of sources can dominate the cosmic-ray spectrum around the knee. X-ray luminosities of Galactic BH binaries in the quiescent state are far below the Eddington luminosity, and thus, radiatively inefficient accretion flows (RIAFs) are formed in the inner region. Strong thermal and turbulent pressures in RIAFs produce outflows, which can create large-scale poloidal magnetic fields. These fields are carried to the vicinity of the BH by the rapid inflow motion, forming a MAD. Inside the MAD, non-thermal protons and electrons are naturally accelerated by magnetic reconnections or stochastic acceleration by turbulence. Both thermal and non-thermal electrons emit broadband photons via synchrotron emission, which are broadly consistent with the optical and X-ray data of the quiescent BH X-ray binaries. Moreover, protons are accelerated up to PeV energies and diffusively escape from these MADs, which can account for the cosmic-ray intensity around the knee energy.

\end{abstract}

\keywords{Stellar mass black holes (1611), Low-mass x-ray binary stars (939), Accretion (14), Non-thermal radiation sources (1119), Cosmic ray sources (328)}


%
\section{Introduction}

Stellar mass black holes (BHs) in close binaries can emit X-rays through mass accretion from their companion stars. Some emit bright X-rays persistently, whereas the majority are discovered as transient sources \citep{2016ApJS..222...15T,2016A&A...587A..61C}. They spend most of their lifetime in the quiescent state, where their luminosities are far below the Eddington luminosity.

Quiescent states of BH X-ray binaries are detected in the radio, infrared/optical, and X-ray bands. The radio emission is widely believed to originate from jets. On the other hand, the origins of the infrared/optical and X-ray signals are still controversial. Hot accretion flow models have been actively discussed since late 90s \citep{EMN97a,1997ApJ...482..448N}. 
However, Chandra and XMM-Newton data indicate that the quiescent spectra are well described by a simple power-law \citep{2008MNRAS.389..423P}, which is in tension with the bumpy spectra predicted by hot accretion flow models \citep{mmk97,1999ApJ...520..298Q}.
Jet dominated models are also proposed, where non-thermal electrons accelerated in the jets provide the radio and X-ray signals via synchrotron radiation \citep{2003MNRAS.343L..99F,2008MNRAS.389..423P,2015MNRAS.446.4098P}. However, these models often require additional components, such as an outer optically thick accretion disk or another emission zone in the jet, to explain the infrared and optical data.

In this paper, we propose magnetically arrested disks (MADs) in quiescent BH X-ray binaries (QBXBs) as the origin of the infrared/optical and X-ray emission. A MAD is a strongly magnetized accretion flow where the magnetic fields affect its dynamics. In MADs, the magnetic energy dissipates through magnetic reconnections \citep{2017ApJ...850...29R,2018ApJ...868L..18H} and/or turbulent cascades \citep{how10,2019PNAS..116..771K}, which heat up thermal electrons to relativistic energies. They emit infrared/optical photons via cyclosynchrotron radiation. Non-thermal electrons are also accelerated in MADs by magnetic recconections \citep{2001ApJ...562L..63Z,2014ApJ...783L..21S} and/or stochastic acceleration by turbulence \citep{KTS16a,2019MNRAS.485..163K,2018PhRvL.121y5101C,2018ApJ...867L..18Z}, and they emit power-law X-rays via synchrotron radiation. MADs can launch a relativistic jet \citep{TNM11a,MTB12a}, which is responsible for the radio data. It is believed that MADs are formed in radio galaxies, which is supported by magnetic field estimates based on radio observations \citep{2014Natur.510..126Z,2015MNRAS.451..927Z}. However, the formation mechanism of MADs are not well established yet (see \citealt{2020ApJ...896L...6R} for the case of Sgr A*).

MADs can also accelerate non-thermal protons, or cosmic-ray (CR) protons, which can diffusively escape from the MADs. 
CRs of energies below PeV  (``knee'') are believed to be produced in the Milky Way. Supernova remnants are considered to produce Galactic CRs through diffusive shock acceleration \citep[see][for reviews]{2012SSRv..173..491S,2012SSRv..173..369H}, and their gamma-ray spectra show that they authentically accelerate protons to GeV - TeV energies \citep{Ackermann:2013wqa,Abeysekara:2020uqf}. However, they often show spectral cutoffs or breaks below $\sim10$ TeV \citep{Aha13a,Abeysekara:2020uqf}, calling into question whether they are capable of generating PeV-scale particles\footnote{Recently, \cite{2020ApJ...896L..29A} and \cite{2021NatAs.tmp...41T} independently reported detection of $\gtrsim100$ TeV photons from a SNR, G106.3+2.7. These findings suggest it as a potential PeVatron.}.
This raised the need for exploring other candidates, which include Sgr A* \citep{2016Natur.531..476H,FMK17a}, milli-second pulsars \citep{2018JCAP...07..042G}, isolated black holes (BHs; \citealt{IMT17a}), jets in X-ray binaries \citep{2020MNRAS.493.3212C}, pulsar wind nebulae \citep{2018MNRAS.478..926O}, stellar winds from young star clusters \citep{Aharonian:2018oau}, and superbubbles \citep{Bykov:2014asa}.
Very recently, Tibet AS$\gamma$ Collaboration reported discovery of diffuse sub-PeV gamma-rays from the Galactic plane, proving that PeVatrons exist in our Galaxy~\citep{2021PhRvL.126n1101A}. Following multi-messenger discussions suggest that PeVatrons can be a population distinct from GeV-TeV CR sources~\citep{2021arXiv210405609L,2021arXiv210409491F}, which strengthens the need for other PeVatron candidates.
In this paper, we newly add MADs in QBXB (QBXB-MADs) into the list.

This paper is organized as follows. In Section \ref{sec:MAD}, we discuss conditions for the MAD formation in stellar-mass BH binaries and demonstrate that they are fulfilled in the quiescent state. In Section \ref{sec:photon}, we study emission from thermal and non-thermal electrons in the QBXB-MADs. We focus on a few selected QBXBs that have well-measured multiwavelength spectra. We then show that our QBXB-MAD model is in reasonable agreement with the observed data, which supports our assertion that QBXBs form MADs. In Section \ref{sec:CR}, we examine the production of CRs in QBXB-MADs. We demonstrate that they can produce PeV-scale protons, potentially dominating the observed CR spectrum around the knee. In Section \ref{sec:discussion}, we discuss implications and outline strategies to test our key assumptions. In Section \ref{sec:summary}, we present our conclusions. We use the convention of $Q_x = Q/10^x$ in cgs units unless otherwise noted.

\section{Realization of QBXB-MADs}\label{sec:MAD}

  \begin{figure}
   \begin{center}
    \includegraphics[width=\linewidth]{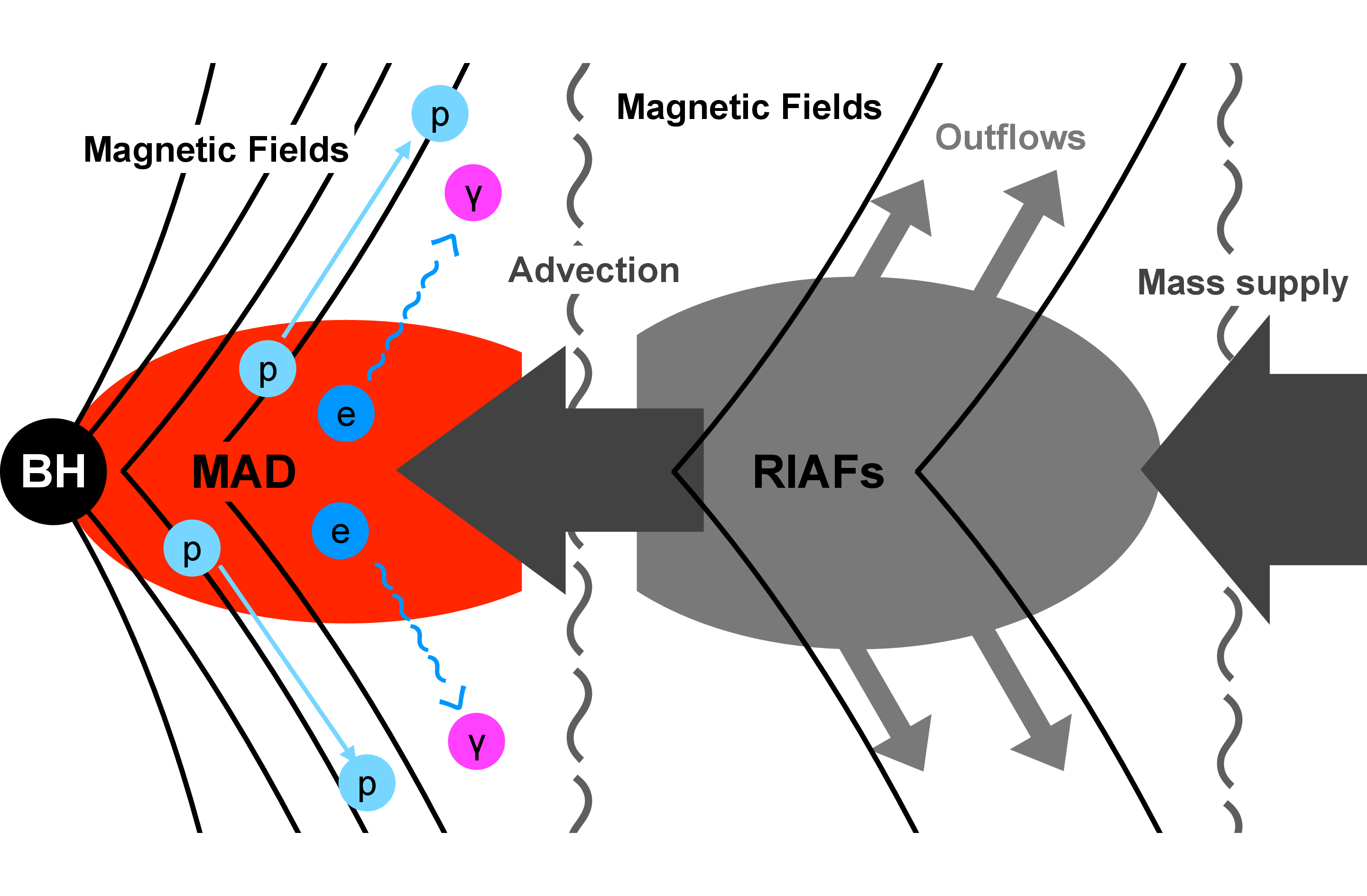}
    \caption{Schematic picture of our QBXB-MAD scenario. In the quiescent state, the standard optically thick disk is truncated at the outer part of the accretion flow. Inside the truncation radius, the accretion flow is in a radiatively inefficient state where outflows are produced. These outflows stretch the magnetic field generated by magneto-rotational instability (MRI) and shear motion, making large-scale poloidal fields. The poloidal fields are advected toward the vicinity of the BH, which results in formation of a MAD. Magnetic reconnections directly heat up thermal electrons and accelerate CR electrons, leading to efficient synchrotron emission that can account for optical and X-ray data. CR protons are also accelerated, and they diffusively escape from the system without losing their energies, possibly providing a dominant contribution to the observed intensity of PeV CRs. }
    \label{fig:schematic}
   \end{center}
  \end{figure}

Figure \ref{fig:schematic} shows the schematic picture of our scenario. In QBXBs, the mass accretion rate is so low that the accretion flows cannot cool through radiative processes. Then, the optically thick accretion disk should be truncated at an outer radius, and a radiatively inefficient accretion flow (RIAF; \citealt{ny94,YN14a}) is formed inside the truncation radius \citep{EMN97a}. The accretion flow is turbulent due to magneto-rotational instability (MRI; \citealt{bh91}), and the turbulent viscosity and magnetic torque enable a steady accretion. Thermal, magnetic, and turbulence pressures drive outflows as seen in magnetohydrodynamic (MHD) simulations \citep{OM11a,SNP13a,2015ApJ...804..101Y,2019ApJ...875L...5E}. These outflows convert the toroidal magnetic fields generated by the shear motion to the poloidal fields \citep{2020MNRAS.494.3656L}. The rapid infall motion of the RIAFs can carry these poloidal fields to the inner region. Then, the magnetic flux is accumulated at the vicinity of the BH, leading to the formation of a MAD~\citep{2011ApJ...737...94C}. In this section, we discuss the feasibility of our QBXB-MAD scenario based on the current understanding of the plasma and accretion physics. We define a MAD as an accretion flow with $\beta\lesssim1$, where $\beta$ is the plasma beta. Here, we ignore the magnetic flux carried from the companion star. If we take it into account, MADs are more likely to be formed. In this sense, our estimates in this section is conservative.

In RIAFs, the matter cannot cool within the infall timescale, which results in a proton temperature comparable to the virial temperature. RIAFs are geometrically thick because of the strong thermal pressure. The thick geometry allows a large turbulent eddy, which leads to a large turbulent viscosity. Then, the angular momentum transport is efficient, resulting in a radial motion faster than the standard thin disk. 
Since RIAFs produce outflows, the mass accretion rate can depend on the distance from the BH, $R$, and written as $\dot{M}(R)=(R/R_{\rm trn})^{s_w}\dot{M}_o$, where $R_{\rm trn}$ is the truncation radius, $\dot{M}_o$ is the mass accretion rate at $R=R_{\rm trn}$, and $s_w$ is a parameter that describes the outflow efficiency \citep{BB99a}. 
The radial velocity, sound velocity, and density in RIAFs can be analytically estimated to be \citep[see][for parameter sets for active galactic nuclei (AGN)]{2019PhRvD.100h3014K,Kimura:2020thg} 
\begin{eqnarray}
& &V_R\approx \frac12\alpha V_K\simeq4.7\times10^7\alpha_{-0.5}\mathcal{R}_4^{-1/2}\rm~cm~s^{-1},\\
& &C_s\approx \frac12V_K\simeq1.5\times10^8\mathcal{R}_4^{-1/2} \rm~cm~s^{-1},\\
& &N_p\approx \frac{\dot{M}(R)}{4\pi RHV_Rm_p}\\
& &\simeq1.3\times10^{11}\left(\frac{\dot{m}(R)}{0.01}\right)M_1^{-1}\mathcal{R}_4^{-3/2}\alpha_{-0.5}^{-1}\rm~cm^{-3},\nonumber
\end{eqnarray}
where $M$ is the BH mass, $M_1=M/(10 M_\odot)$, $V_K=\sqrt{GM/R}$ is the Keplerian velocity, $\alpha$ is the viscous parameter \citep{ss73}, $H\approx (C_s/V_K)R\approx R/2$ is the scale height, $\dot{m}(R)=\dot{M}(R)c^2/L_{\rm Edd}$, $\mathcal{R}=R/R_G$, $L_{\rm Edd}$ is the Eddington luminosity, $m_p$ is the proton mass, and $R_G=GM/c^2$ is the gravitational radius. The pre-factors in $V_R$ and $C_s$ are determined so that the quantities are consistent with recent MHD simulations \citep{OM11a,2019MNRAS.485..163K}.

Electrons and protons in RIAFs are thermally decoupled because of a long relaxation timescale. Electrons in collisionless plasma can receive a significant fraction of the dissipation energy by magnetic reconnections and turbulence cascades \citep{2017ApJ...850...29R,2019PNAS..116..771K}, and they do not efficiently cool if they are non-relativistic. Thus, electrons are expected to be close to the virial temperatures for $R\gtrsim (m_p/m_e)R_G$, where $m_e$ is the electron mass. On the other hand, electrons become relativistic for the inner region and efficiently cool via synchrotron and Comptonization processes \citep[e.g.,][]{NYM95a,mmk97,kmt15}. Then, the electron temperature is almost independent of $\mathcal{R}$ and frozen to the mildly relativistic regime \citep[e.g.,][]{ny95,Kimura:2020thg}.

The truncation radius can be estimated by balancing the accretion timescale and proton cooling timescales. The accretion timescale for a RIAF can be estimated to be 
\begin{equation}
t_{\rm fall}\approx\frac{R}{V_R}\simeq3.1\times10^2\mathcal{R}_4^{3/2}M_1\alpha_{-0.5}^{-1}{\rm~sec}.
\end{equation}
Thermal protons mainly lose their energies via Coulomb scattering with electrons. The energy loss timescale is given by
\begin{equation}
 t_{pe} = \frac{\sqrt{2\pi}}{2N_p\sigma_Tc\ln\Lambda}\frac{m_p}{m_e}\theta_e^{3/2},\label{eq:tpe}
\end{equation}
where $\sigma_T$ is the Thomson cross section, $\ln\Lambda\sim20$ is the Coulomb logarithm, and $\theta_e=k_BT_e/(m_ec^2)$ is the normalized electron temperature. To obtain Equation (\ref{eq:tpe}), we assume non-relativistic limit and $\theta_e>\theta_p$, where $\theta_p=k_BT_p/(m_pc^2)$ is the normalized proton temperature. The latter condition is always satisfied for RIAFs, while electrons can be relativistic at $R\lesssim (m_p/m_e)R_G$.
The cooling timescale of thermal electrons is shorter than the Coulomb interaction timescales in RIAFs, and thus, Equation (\ref{eq:tpe}) is regarded as the cooling timescale of RIAF plasma. In cases with a sufficiently low $\dot{m}$, the temperatures in the outer region can be approximated as $kT_e\sim kT_p\sim m_pC_s^2$, i.e., $\theta_e\approx(m_p/m_e)(V_K/c)^2/4$, which leads to 
\begin{equation}
 t_{pe}\simeq4.4\times10^2 \left(\frac{\dot{m}(R)}{0.01}\right)^{-1}M_1\alpha_{-0.5}\rm~sec.
\end{equation}
Equating $t_{pe}$ and $t_{\rm fall}$, we estimate the truncation radius to be
\begin{equation}
 \mathcal{R}_{\rm trn}\approx1.3\times10^4\alpha_{-0.5}^{4/3}\dot{m}_{o,-2}^{-2/3},\label{eq:Rtrn}
\end{equation}
where $\dot{m}_o=\dot{M}_oc^2/L_{\rm Edd}$.
Therefore, the truncation radius can be larger than $(m_p/m_e)R_G$ when $\dot{m}_o\lesssim0.1\alpha_{-0.5}^2$, which justifies our assumption of $kT_e\sim kT_p$. Such a large truncation radius is supported by the spectroscopic observation of QBXBs. The widths of double-peak emission lines in the quiescent state indicate a circular motion of $\sim10^3\rm~km~s^{-1}$, which infer a thin disk truncated at $R_{\rm trn}\sim10^4R_G$ \citep{1994ApJ...436..848O,2003ApJ...593..435M}.

In RIAFs, strong poloidal fields can be generated even from a small seed field. We estimate the poloidal field strength at the wind launching region. Non-liner growth of MRI creates the magnetic field of $\beta\approx 8\pi m_pN_pC_s^2/B^2 \sim10-100$, which is dominated by the toroidal field \citep{HRG13a,2019MNRAS.485..163K}. Thermal and magnetic pressures drive outflows \citep{OM11a,SNP13a,2015ApJ...804..101Y,2019ApJ...875L...5E}, which convert the toroidal magnetic fields to the poloidal fields. The poloidal magnetic field is as strong as 30-40\% of the total magnetic field  based on MHD simulations \citep{HRG13a,KTS16a,2019MNRAS.485..163K}, although these values contain the turbulent component. Assuming that the global poloidal field strength is comparable to the turbulent poloidal fields, we estimate the plasma beta by the global poloidal fields to be $\beta_p= 8\pi m_pN_pC_s^2/B_p^2 \sim10^3-10^4$, where $B_p$ is the poloidal field strength.

The rapid infall motion of RIAFs can carry these poloidal fields to the inner region. If we assume the magnetic flux freezing, the poloidal field and gas pressure scale as $B_p\propto R^{-2}$ and $N_pC_s^2\propto R^{-5/2+s_w}$, respectively. Thus, the poloidal plasma beta scales as $R^{3/2+s_w}$. 
Then, the accreting plasma of $\beta_{p,o}$ at $R\sim R_{\rm trn}$ would become $\beta_{p,i}\sim (R_G/R_{\rm trn})^{3/2+s_w}\beta_{p,o}$ at $R\sim R_G$. For $\beta_{p,o}\sim10^4$ and $R_{\rm trn}\sim10^4R_G$, we obtain  $\beta_{p,i}\sim10^{-3}$ for $s_w\sim0.25$. We note that $s_w\sim0.2-0.3$ is often used when ones try to fit the broadband spectra in low-Eddington objects \citep{yqn03,nse14}.

Such a strongly magnetized flow is unlikely to be realized, because the strong magnetic fields halt the accretion. In reality, some mechanisms, such as magnetic Rayleigh-Taylor instability (MRTI), should prevent the enhancement of the poloidal field by disturbing the poloidal field locally. This will maintain the moderately strong magnetic field of $\beta\sim0.1-1$, and form a MAD where the magnetic flux threading the horizon is equal to the saturation value \citep{TNM11a,2012MNRAS.426.3241N}. Indeed, analytic modelings of magnetic fluxes suggest that rapid inward motion of RIAFs can result in accumulation of poloidal magnetic fields at the vicinity of the BH \citep{2011ApJ...737...94C}. Also, magnetic flux accumulation by advection is discussed in the context of the jetted tidal disruption event \citep{2014MNRAS.437.2744T}. The high-resolution and long-term general relativistic (GR) MHD simulations support the idea that the outflows create poloidal fields from toroidal fields and a MAD is eventually formed by advection of the poloidal fields \citep{2020MNRAS.494.3656L}. In light of these considerations, we conclude that Galacitc X-ray binaries in their quiescent state very likely host MADs.

We cautiously note that our argument is applicable only for highly sub-Eddington systems, e.g., quiescent states in X-ray binaries. RIAFs with relatively high mass accretion rates ($\dot{m}_o\gtrsim0.01$) are expected in low-hard states, and the truncation radius in this state should be smaller \citep{EMN97a}. In this situation, MADs are unlikely to be formed (see Section \ref{sec:discussion} for detail). Also, some long-term GRMHD simulations do not arrive at the MAD state \citep{2012MNRAS.426.3241N,2020ApJ...891...63W}. This may indicate that our scenario may not be always applicable even for highly sub-Eddington systems. Currently, the exact condition of MAD formation is still unclear, and further investigation is necessary in order to understand whether they are ubiquitous in highly sub-Eddington systems.

\section{Photon spectra from QBXB-MADs}\label{sec:photon}

\begin{table}[t]
\begin{center}
\caption{List of model parameters and physical quantities. The references for BH masses and distances are \cite{2010ApJ...710.1127C,2019MNRAS.485.2642G} for A0620-00, \cite{2010ApJ...716.1105K,2009ApJ...706L.230M} for V404 Cyg, and \cite{2013AJ....145...21K,2006ApJ...642..438G} for XTE J1118+480.}\label{tab:param}
Shared Parameters
 \begin{tabular}{|cccccc|}
\hline
  $\alpha$ & $\mathcal R$ & $\epsilon_{\rm dis}$ & $\eta$ & $\epsilon_{\rm NT}$ & $s_{\rm inj}$ \\
\hline
  0.3  & 10 & 0.15 & 5 &  0.33 & 1.3 \\
\hline
 \end{tabular}

Parameters for individual BHs
 \begin{tabular}{|c|cccc|}
\hline
Name &  $M~[\msun]$ & $\dot{m}~[10^{-4}]$ & $\beta$ &  $d_L~[\rm kpc]$ \\
\hline
A0620-00 & 6.6 &  1.0 & 0.40 & 1.7 \\
V404 Cyg & 9.0 & 2.0 & 0.50 & 2.4 \\
XTE J1118+480 & 7.5 & 0.10 & 0.10 & 1.7 \\
Hypothetical\footnote{Cases shown in Figure \ref{fig:typ}, Figure \ref{fig:Loir}, and Figure \ref{fig:CR}} & 10 & $0.1-100$ & $0.1-0.5$ & $2-8$\\
\hline
 \end{tabular}
\end{center}
\end{table}

MADs dissipate their magnetic energy through magnetic reconnections and/or turbulence cascades, which heat up the thermal plasma and accelerate CR protons and electrons. 
We calculate the photon spectra from the QBXB-MADs based on the formalism in \cite{2020ApJ...905..178K}, where one-zone and steady state approximations are utilized. We consider an accreting plasma\footnote{We consider accretion rate at $R=\mathcal{R}R_G$, which should be lower than $\dot{M}_o$ in Section \ref{sec:MAD} because $\dot{M}(R)\approx\dot{M}_o(R/R_{\rm trn})^{s_w}$ with $s_w>0$. $s_w$, $R_{\rm trn}$, and $\dot{M}_o$ do not appear explicitly in our calculations of the emission from the MAD.} of size $R=\mathcal{R}R_G=\mathcal{R}GM/c^2\simeq1.5\times10^7\mathcal{R}_1M_1$ cm with the mass accretion rate $\dot{M}=\dot{m}L_{\rm Edd}/c^2\simeq2.2\times10^{-12}~M_\odot~{\rm yr^{-1}}~\dot{m}_{-4}M_1$.
To estimate physical quantities in MADs, we use the analytic prescription for RIAFs with a parameter set appropriate for MADs. The radial velocity and magnetic field of the QBXB-MAD are written as 
\begin{eqnarray}
& & V_R\simeq1.5\times10^9\mathcal{R}_1^{-1/2}\alpha_{-0.5}\rm~cm~s^{-1}~\\
& &B\approx \sqrt{\frac{8\pi m_pN_pC_s^2}{\beta}}\label{eq:BinMAD}\\
& &\simeq6.2\times10^5\mathcal{R}_1^{-5/4}\dot{m}_{-4}^{1/2}M_1^{-1/2}\alpha_{-0.5}^{-1/2}\beta_{-1}^{-1/2}\rm~G,\nonumber
\end{eqnarray}
 where we provide $\beta$ as a parameter. These analytic expressions are in rough agreement with the results of GR MHD simulations of MADs \citep{2012MNRAS.426.3241N,2019ApJ...874..168W}.

The energy dissipation rate can be parameterized as $Q_{\rm dis}\approx\epsilon_{\rm dis}\dot{m}L_{\rm Edd}$, where $\epsilon_{\rm dis}$ is the dissipation parameter.  Based on PIC simulations, the energy partition between protons and electrons are given by $Q_p/Q_e\approx(m_p/m_e)^{1/4}(T_p/T_e)^{1/4}$ \citep{2018ApJ...868L..18H}. Introducing the non-thermal particle production efficiency, $\epsilon_{\rm NT}$, we write the CR proton and electron luminosities as $L_{p,\rm CR} = \epsilon_{\rm NT}\epsilon_{\rm dis}\dot{m}L_{\rm Edd}$ and $L_{e,\rm CR}=\epsilon_{\rm NT}(Q_e/Q_p)\epsilon_{\rm dis}\dot{m}L_{\rm Edd}$, respectively. The heating rate of thermal electrons is then given by $L_{e,\rm thrml} = (1-\epsilon_{\rm NT})(Q_e/Q_p)\epsilon_{\rm dis}\dot{m}L_{\rm Edd}$.  The proton temperature can be estimated as $k_BT_p \approx m_pC_s^2$ (see the next paragraph for the electron temperature). We consider 4 components: thermal electrons, CR electrons, CR protons, and secondary electron-positron pairs. 

First, we describe emissions from the thermal component. The thermal electrons emit photons via cyclosynchrotron, bremsstrahlung, and Comptonization processes, whose emissivities strongly depend on the electron temperature. We calculate the photon spectra by the method given in Appendix of \cite{kmt15}, except for the fitting formula of the thermal cyclosynchrotron emission; we utilize a formula for sub- and mildly relativistic electrons given in \cite{1996ApJ...465..327M}. We compute the cooling rate by integrating the emitted photon spectrum, and obtain the electron temperature by balancing the cooling rate with the heating rate iteratively.

Next, we explain the CR components. We solve the transport equations for the CR protons and electrons:
\begin{equation}
 -\frac{d}{dE_i}\left(\frac{E_iN_{E_i}}{t_{i,\rm cool}}\right)=-\frac{N_{E_i}}{t_{\rm esc}}+\dot{N}_{E_i,\rm inj},
\end{equation}
where $i=e,p$ indicates the particle species, $E_i$ is the particle energy, $N_{E_i}$ is the number spectrum, $\dot{N}_{E_i,\rm inj}$ is the injection term, and $t_{i,\rm cool}$ and $t_{\rm esc}$\footnote{The value of $t_{\rm esc}$ for CR protons and CR electrons for a given energy are the same, so we omit $i$ from $t_{\rm esc}$.} are the cooling and escape timescales, respectively. We use a single power-law injection function with an index $s_{\rm inj}$ and exponential cutoff at $E_{i,\rm cut}$, i.e., $\dot{N}_{E_i,\rm inj} \propto (E_i/E_{i,\rm cut})^{-s_{\rm inj}}\exp(-E_i/E_{i,\rm cut})$, normalizing it by $L_{i,\rm CR}=\int\dot{N}_{E_i,\rm inj}E_idE_i$.

The cutoff energy is obtained by balancing the acceleration and loss timescales. The acceleration timescale is given by $t_{i,\rm acc}\approx\eta{r_{i,L}}/(c\beta_A^2)$, where $\beta_A=B/\sqrt{4\pi{N_p}m_pc^2}$ is the Alfven velocity in unit of $c$, $\eta{r_{i,L}}$ is the particle mean-free path, $r_{i,L}=E_i/(eB)$ is the Larmor radius, and $\eta$ is the acceleration efficiency parameter. 
The loss timescale is $t_{i,\rm loss}^{-1}=t_{i,\rm cool}^{-1}+t_{\rm esc}^{-1}$. We consider advective (infall to the BH) and diffusive escapes. The advective and diffusive escape timescales are given by $t_{\rm adv}=R/V_R$ and $t_{\rm diff}=3R^2/(\eta{r_{i,L}c})$, respectively, and the escape timescale is given by $t_{\rm esc}^{-1}=t_{\rm adv}^{-1}+t_{\rm diff}^{-1}$.
As the cooling process of CR electrons, we only consider synchrotron emission, $t_{e,\rm cool}=t_{e,\rm syn}$, where $t_{e,\rm syn}$ is the synchrotron cooling timescale for electrons, because other processes are negligible\footnote{We can ignore inverse Compton scattering and bremsstrahlung emissions by CR electrons in QBXB-MADs. The magnetic field energy density is higher than the photon energy density by about two orders of magnitude, which enables us to ignore the inverse Compton emission. The plasma density in the QBXB-MAD is so low that the bremsstrahlung emission provides a negligible contribution.}.
Regarding the CR protons, we consider $pp$ inelastic collisions ($p+p\rightarrow{p}+p(n)+\pi$), photomeson production ($p+\gamma\rightarrow{p}(n)+\pi$), Bethe-Heitler ($p+\gamma\rightarrow p+e^++e^-$), and proton synchrotron processes. We write the proton cooling timescale as $t_{p,\rm cool}^{-1}=t_{pp}^{-1}+t_{p\gamma}^{-1}+t_{\rm BH}^{-1}+t_{p,\rm syn}^{-1}$, where $t_{pp}$, $t_{p\gamma}$, $t_{\rm BH}$, and $t_{p,\rm syn}$ are the cooling timescales by $pp$ inelastic collision, photomeson production, Bethe-Heitler, and proton synchrotoron processes, respectively. The expressions for them are given in \cite{2019PhRvD.100h3014K}, and we appropriately take into account the energy dependent cross sections \citep{SG83a,CZS92a,MN06b,2014PhRvD..90l3014K}. 
In the range of our interest, the proton synchrotron process dominates over the other three\footnote{$pp$ inelastic collisions and photomeson production processes can produce high-energy neutrinos, and thus, these are relevant in view of multi-messenger astrophysics. However, we found that QBXB-MADs are too faint to be detected by high-energy neutrino telescopes near future.}.
 The secondary electron-positron pairs are produced via Bethe-Heitler processes, decay of pions, and two-photon pair productions. They emit gamma-rays via synchrotron emission. Although we approximately take into account the emission by secondary electron-positron pairs, their contributions to the resulting photon spectra are sub-dominant in the range of our interest.

  \begin{figure}[tbp]
   \begin{center}
    \includegraphics[width=\linewidth]{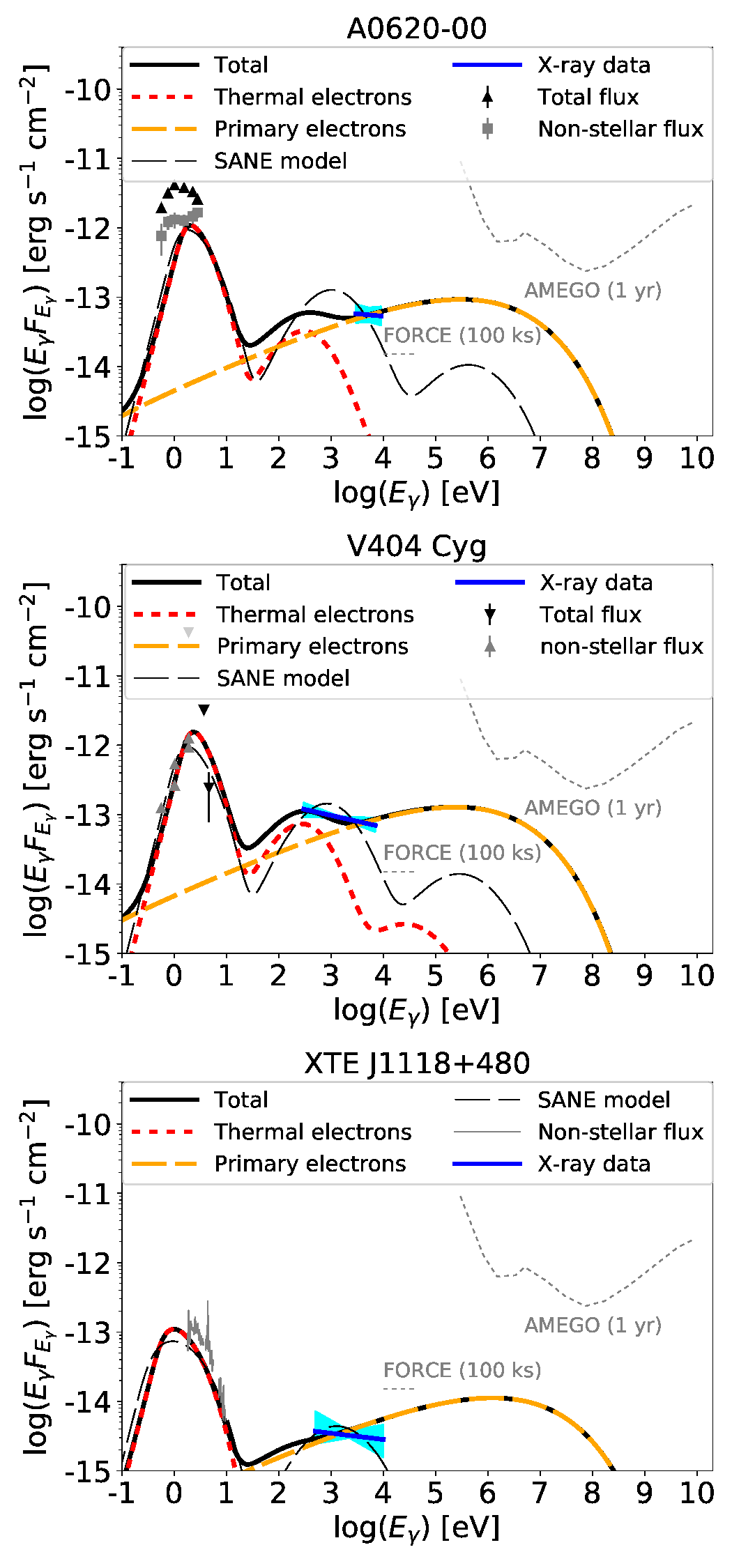}     
    \caption{Broadband spectra for well-known BH X-ray binaries, A0620-00 (top), V404 Cyg (middle), and XTE J1118-480 (bottom), in quiescent states. Thick lines are photon spectra by the MAD scenario (this study) and the thin-dashed lines are ones for a weak magnetic field scenario (SANE; see Section \ref{sec:discussion}). The total fluxes of the optical band are given by black points, and the gray points indicate non-stellar fluxes. The blue shaded regions show the power-law fit of X-ray data. The thin dotted lines show the sensitivity curves for FORCE (100 ks; \citealt{2018SPIE10699E..2DN}) and AMEGO (1-year; \citealt{2017ICRC...35..798M}).  The data are taken from \cite{2018ApJ...852....4D} for A0620-00, \cite{2004MNRAS.352..877Z,2009MNRAS.399.2239H} for V404 Cyg, and \cite{2003ApJ...593..435M,2013ApJ...773...59P} for XTE J1118+480. 
}
    \label{fig:spe}
   \end{center}
  \end{figure}

  \begin{figure}[tbp]
   \begin{center}
    \includegraphics[width=\linewidth]{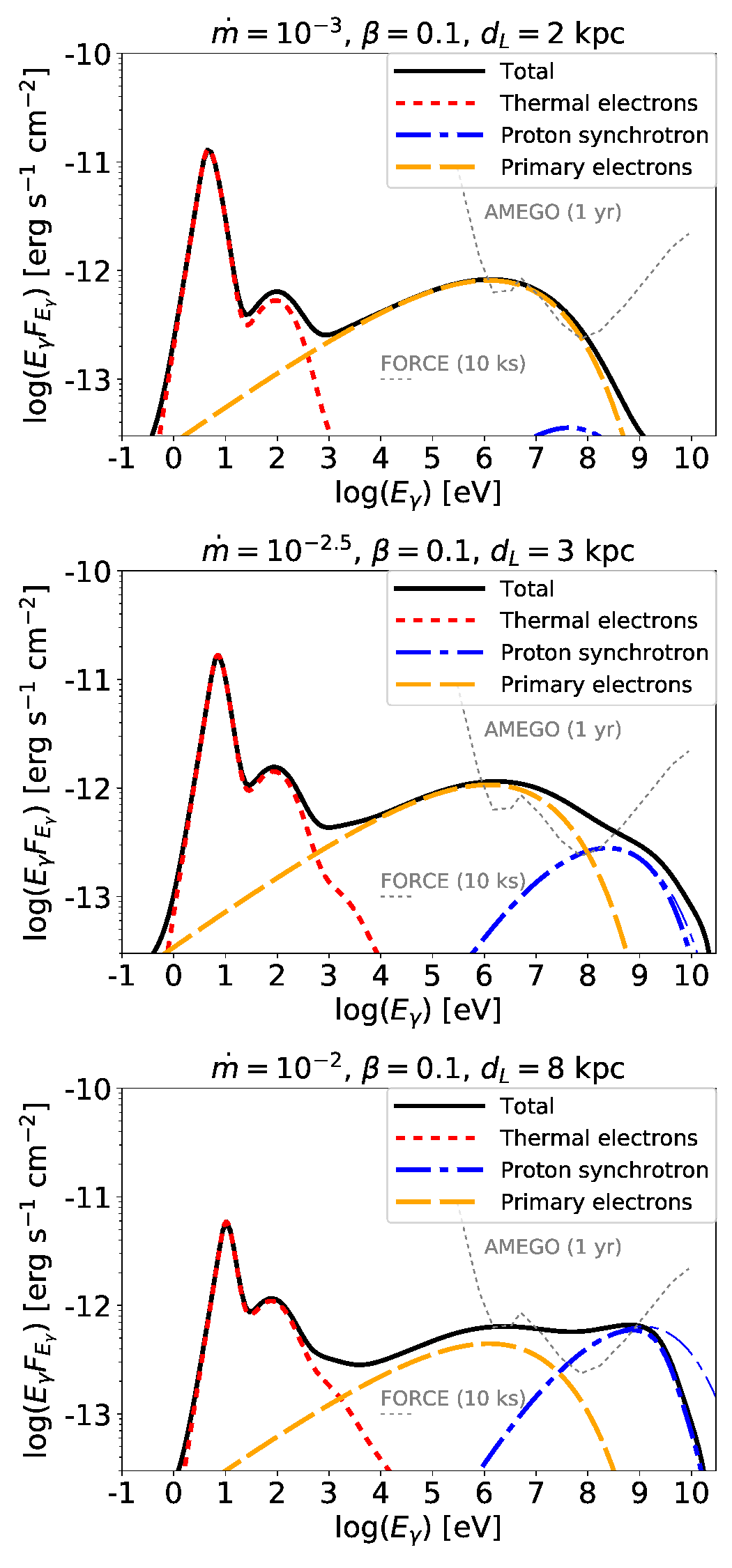}     
    \caption{Same as Figure \ref{fig:spe}, but for hypothetical X-ray binaries with higher values of $\dot{m}$. The integration time for FORCE sensitivity curves are also changed to 10 ks. In the bottom panel, the thin line represents the intrinsic spectrum at the source, and the thick line is the flux attenuated by $\gamma\gamma$ pair production.
}
    \label{fig:typ}
   \end{center}
  \end{figure}

  \begin{figure}[t]
   \begin{center}
    \includegraphics[width=\linewidth]{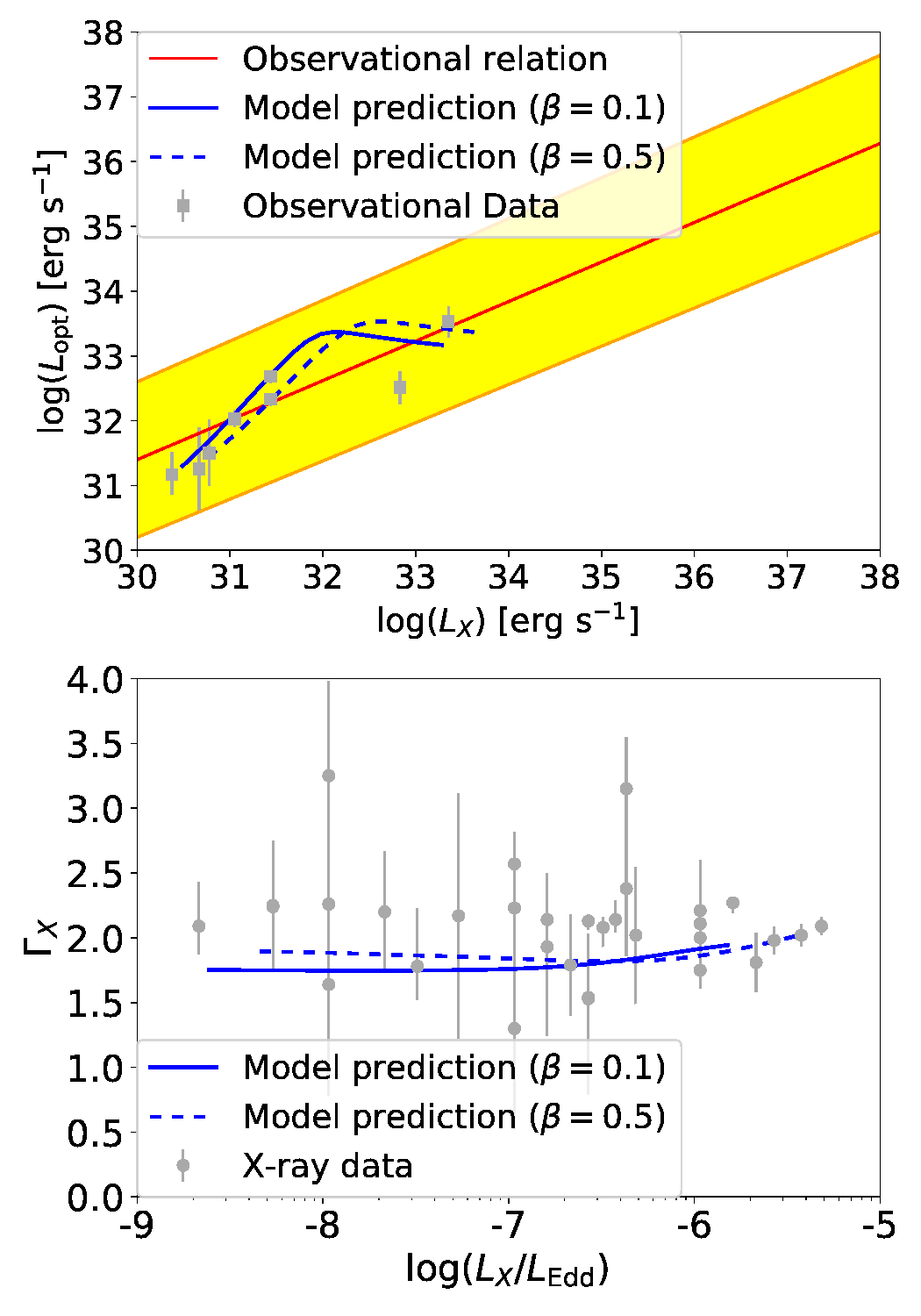}
    \caption{Top panel: $L_{\rm opt}-L_X$ relation for our QBXB-MAD scenario  (solid line) and observations of QBXBs (data points; \citealt{2006MNRAS.371.1334R}). The yellow regions indicates the observed $L_{\rm opt}-L_X$ relation including both quiescent and hard states \citep{2006MNRAS.371.1334R}.
    Bottom panel: $\Gamma_X-L_X/L_{\rm Edd}$ relation in our QBXB-MAD scenario (solid line) and observations (data points; \citealt{2013ApJ...773...59P}).}
    \label{fig:Loir}
   \end{center}
  \end{figure}

Figure \ref{fig:spe} indicates broadband photon spectra from the QBXB-MADs for three BH X-ray binaries, whose parameters are given in Table \ref{tab:param}. We select these objects because the qualities of their data sets in the quiescent states are better than others.
Our scenario is in broad agreement with the optical/infrared and X-ray data. Thermal electrons emit optical photons by cyclosynchrotron emission. We find that MADs with $\dot{m}\gtrsim10^{-5}$ are still optically thick for synchrotron-self absorption at the synchrotron characteristic frequency. In this case, the peak frequency of the synchrotron spectrum is estimated to be \citep{1997ApJ...477..585M,Kimura:2020thg}
\begin{equation}
E_{\rm syn,abs}\approx\frac{3x_MeB\theta_e^2h_p}{4\pi{m_e}c}\simeq1.1B_{5.5}\theta_e^2\left(\frac{x_M}{200}\right)\rm~eV,
\end{equation}
where $x_M=E_{\rm syn,abs}/E_{\rm syn,pk}\approx2.2\times10^3\dot{m}^{1/4}$ is the conversion factor from the synchrotron frequency,  $E_{\rm syn,pk}=3\theta_e^2h_peB/(4\pi{m_e}c)$, to the spectral peak, $E_{\rm syn,abs}$ \citep{1997ApJ...477..585M}, and $h_p$ is the Planck constant. This value is in agreement with the peak energy of the optical non-stellar component. The Thomson optical depth of QBXB-MADs is too low to emit X-rays by Comptonization, but CR electrons in QBXB-MADs can emit high-energy photons up to the MeV range. Equating $t_{e,\rm cool}$ to $t_{\rm fall}$ and $t_{\rm acc}$, we estimate the cooling and cutoff energies to be $E_{\gamma,\rm cl}\approx h_peB\gamma_{e,\rm cl}^2/(2\pi m_ec)\sim3.7\times10^{-3}B_{5.5}\gamma_{e,\rm cl}^2$ eV and $E_{\gamma,\rm cut}\approx3e^2h_p\beta_A^2/(m_ec\sigma_T\eta)\sim15(\beta_A/0.7)^2(\eta/5)^{-1}$ MeV, respectively, where $\gamma_{e,\rm cl}\approx {\rm max}(1,~6\pi m_ecV_R/(\sigma_TB^2R))$ is the electron Lorentz factor at the cooling break. In our parameter choice, $\gamma_{e,\rm cl}=1$ is realized. The X-ray band lies between the two frequencies where the synchrotron spectrum is given by $E_\gamma{L}_{E_\gamma}\propto{E}_\gamma^{(2-s_{\rm inj})/2}$. The resulting spectra can reproduce the X-ray data of QBXBs within their uncertainty. Future hard X-ray missions, such as FORCE \citep{2018SPIE10699E..2DN}, will be able to measure the spectrum above 10 keV, which provides a good test of the QBXB-MAD model. Although the spectrum extends to MeV energies, it is too faint to be detected by near future projects, such as e-ASTROGAM \citep{2017ExA....44...25D}, AMEGO~\citep{2017ICRC...35..798M}, and GRAMS \citep{2020APh...114..107A}. 

The QBXBs in Figure \ref{fig:spe} have relatively low mass accretion rates, $\dot{m}\simeq 10^{-5}-2\times10^{-4}$. In our scenario, MADs can be formed in a system with a higher mass accretion rate of $\dot{m}\lesssim10^{-2}$ (see below for an observational support). In Figure \ref{fig:typ}, we show the broadband spectra of hypothetical sources with $10^{-3} \leq \dot{m} \leq 10^{-2}$. The gamma-ray counterpart can be detectable by the MeV satellites up to a few kpc for $\dot{m}\sim10^{-3}$ and close to 10 kpc for $\dot{m}\sim10^{-2}$. Such systems might be discovered by current and future X-ray monitoring satellites. Also, known sources may emit detectable MeV gamma-rays during a specific epoch of an outburst, although the time window for the suitable accretion rate is limited.

The multiwavelength observations of QBXBs show the correlation between the optical luminosity, $L_{\rm opt}$, and X-ray luminosity in the 2-10 keV band, $L_X$ \citep{2006MNRAS.371.1334R}. In addition, the X-ray observations exhibit that the X-ray photon index, $\Gamma_X$, is almost constant for a wide range of the X-ray Eddington ratio of $L_X/L_{\rm Edd}\lesssim5\times10^{-6}$~\citep{2013ApJ...773...59P}\footnote{In \cite{2013ApJ...773...59P}, $L_X$ is defined in the 0.5-10 keV band, while we use $L_X$ of the 2-10 keV band throughout the paper, which results in a factor of $\sim1.9$ difference in $L_X$. Here, we assume the photon index $\Gamma\simeq2.0$ in the X-ray band.}.
We calculate photon spectra for various $\dot{m}$ with a fixed ($M_{\rm BH}/M_\odot,~\beta)=(10,~0.1)$ and (10, 0.5). The resulting spectra are consistent with the observed relations as shown in Figure \ref{fig:Loir}. These results indicate that the accretion flows in QBXBs are in the MAD regime when $L_X\lesssim3\times10^{33}\rm~erg~s^{-1}$, or $\dot{m}\lesssim0.01$. In contrast, BH binaries of $L_X/L_{\rm Edd}>5\times10^{-6}$, i.e. $\dot{m}\gtrsim0.01$, show an anti-correlation between $\Gamma_X$ and $L_X/L_{\rm Edd}$ \citep{2008ApJ...682..212W}. This transition in the $\Gamma_X-L_X/L_{\rm Edd}$ relation implies that QBXB-MADs no longer exist when $\dot{m}\gtrsim 0.01$.

We note that the free parameters in our calculation are only the mass accretion rate, $\dot{m}$, and plasma beta, $\beta$. We calibrate other parameters ($\alpha$, $\mathcal{R}$, $\epsilon_{\rm dis}$, $s_{\rm inj},~\epsilon_{\rm NT}$, $\eta$) so that the emission from MADs in radio galaxies can reproduce the GeV gamma-ray data observed by Fermi \citep{2020ApJ...905..178K}. We expect that the parameters related to dynamics and non-thermal particle production should be similar in the MADs in radio galaxies and X-ray binaries. 
The particle acceleration in MADs should occur by magnetic reconnections or stochastic acceleration by turbulence. The characteristics of these processes are determined by the magnetization parameter, $\sigma$, and the Alfven velocity, $V_A$. Since the temperature and Alfven velocity are independent of $M$ in the RIAF regime \citep{2019PhRvD.100h3014K}, we expect $\sigma$ and $V_A$ are similar in radio galaxies and X-ray binaries. Therefore, the parameters for non-thermal particle production should also be similar.

\section{CRs from QBXB-MADs}\label{sec:CR}

  \begin{figure}
   \begin{center}
    \includegraphics[width=\linewidth]{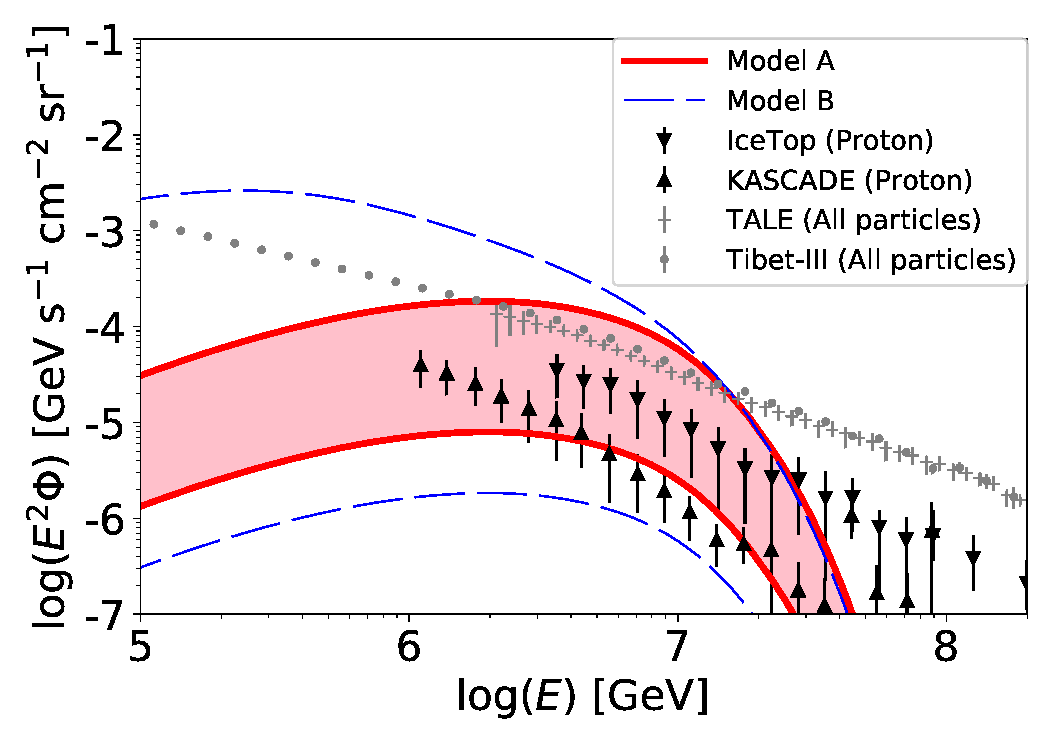}
    \caption{CR spectra predicted by our QBXB-MAD scenario and experimental data. The red and blue lines show the proton CR energy spectra from QBXB-MADs by Method A (population synthesis) and B (X-ray luminosity function), respectively. We use $\beta=0.1$. The uncertainty range by Method A is shown in the pink shaded region. The experimental data for protons and all-particle CR energy spectra are taken from \cite{KASCADE13a,2019PhRvD.100h2002A} and \cite{Amenomori:2008aa,TA18a}, respectively. }
    \label{fig:CR}
   \end{center}
  \end{figure}

Magnetic reconnections in MADs accelerate not only electrons but also protons. CR protons do not have efficient cooling processes in the QBXB-MADs. Significantly high-energy protons can diffusively escape from the system, while lower-energy ones fall to the BH. Equating $t_{\rm fall}$ to $t_{\rm esc}$, we estimate the critical energy of the escaping protons to be $E_{p,\rm esc}\approx3eBV_RR/(c\eta)\simeq19B_{5.5}V_{R,9}R_7(\eta/5)^{-1}$ TeV. The maximum proton energy is given by balancing the escape and acceleration, $E_{p,\rm cut}\approx\sqrt{3}eBR\beta_A/\eta\simeq0.23B_{5.5}R_7(\beta_A/0.7)(\eta/5)^{-1}$ PeV. Thus, QBXB-MADs with $\dot{m}\gtrsim10^{-3}$ can release PeV protons into the interstellar medium (ISM).
Magnetic reconnection or stochastic acceleration processes produce both CR protons and CR heavy nuclei. The abundance ratio in the QBXB-MAD should be similar to the solar abundance ratio, and we can neglect the contribution of CR heavy nuclei if the CR injection efficiency is independent of nuclear species.

In order to calculate the CR intensity, we need to estimate the total number of QBXBs. Since this number is uncertain, we utilize two methods. One (Method A) is based on population synthesis. As an analytic estimate, the number of BH X-ray binaries can be represented as
\begin{eqnarray}
& &N_{\rm BHXB}\sim\rho_{\rm BH}f_{\rm bin}f_{\rm qui}t_{\rm gal}\\
& &\sim3\times10^4\rho_{\rm BH,-2.5}f_{\rm bin,-1.5}f_{\rm qui,-1.5}t_{\rm gal,10},\nonumber
\end{eqnarray}
where $\rho_{\rm BH}$ is the BH formation rate, $f_{\rm bin}$ is the fraction of BHs with a low-mass companion, $f_{\rm qui}$ is the fraction of QBXBs among BH binaries, $t_{\rm gal}$ is the age of our Galaxy, $\rho_{\rm BH,-2.5}=\rho_{\rm BH}/(10^{-2.5} \rm~yr^{-1})$, and $t_{\rm gal,10}=t_{\rm gal}/(10^{10}\rm~yr)$. \footnote{If the typical lifetime of companion stars is shorter than the age of our Galaxy, $t_{\rm gal}$ should be replaced by the lifetime of the companion star. Based on BlackCat \citep{2016A&A...587A..61C}, 15 out of 18 dynamically confirmed BH binaries likely have companions whose lifetimes are longer than the age of the Galaxy, which justifies our estimate.} This crude estimate is roughly consistent with estimates by binary population synthesis models \citep[e.g.,][]{2006A&A...454..559Y}. Independently, $N_{\rm BHXB}\sim10^3$ is suggested by the event rate of the BH X-ray transients \citep{2016A&A...587A..61C} and a recent binary population synthesis model \citep{2020ApJ...898..143S}. Here, we consider that $N_{\rm BHXB}$ ranges from $10^3$ to $3\times10^4$. We assume a flat $\dot{m}$ distribution in logarithmic space in the range of $\dot{m}=10^{-5}-10^{-2}$ for simplicity, although this assumption may be optimistic. QBXB-MADs of $\dot{m}\gtrsim10^{-3}$ may be rarer than those of $\dot{m}\sim10^{-4}$.

The other method (Method B) is based on X-ray luminosity functions. In our QBXB-MAD scenario, the X-ray luminosity is well approximated by $L_X\approx 2.6\times10^{35}\dot{m}\rm~erg~s^{-1}$, and the X-ray luminosity ranges from $3\times10^{30}-3\times10^{33}\rm~erg~s^{-1}$ for $10^{-5}\le\dot{m}\le10^{-2}$. The X-ray luminosity function for $L_X\sim10^{30}-10^{34}\rm~erg~s^{-1}$ is dominated by cataclysmic variables (CVs). The luminosity function for CRs per unit stellar mass is given by $dN/d\log_{10}(L_X)\approx K(L_X/L_b)^{1.22}$, where $K=6.8\times10^{-4}~M_\odot^{-1}$ and $L_b=1.9\times10^{30}\rm~erg~s^{-1}$ \citep{2006A&A...450..117S}. We use the Galactic stellar mass of $M_*=6\times10^{10}~M_\odot$ \citep{2015ApJ...806...96L} to obtain the total number of CVs in the Milky Way. The RXTE survey identified 24 CVs while 21 objects are unidentified. If all the unidentified sources are QBXB-MADs, the X-ray luminosity function of QBXB-MADs can be as high as 87.5\% of that of CVs. This case is regarded as the most optimistic case. On the other hand, the luminosity function of Galactic LMXBs is flat, $dN/d\log_{10}(L_X)\approx 100$, for $10^{35}\rm~erg~s^{-1}<L_X<10^{37}\rm~erg~s^{-1}$  \citep{2006A&A...450..117S}. As the most pessimistic case, we use the extrapolation of the luminosity function of LMXBs toward lower luminosities.

The differential CR-proton injection rate to the ISM is written as 
\begin{equation}
 E_pQ_{E_p}\approx\int\frac{E_p^2N_{E_p}}{t_{\rm diff}}\frac{dN_{\rm LMBH}}{d\dot{m}}d\dot{m}.
\end{equation}
CR protons propagate in the ISM and arrive on Earth. The confinement time in the ISM, $t_{\rm conf}$, can be provided by the grammage, $X_{\rm esc}=n_{\rm ISM}\mu{m_p}ct_{\rm conf}$, where $n_{\rm ISM}$ and $\mu$ are the number density and mean atomic mass of the ISM gas, respectively. Based on the measurements of the boron-to-carbon ratio, the grammage is estimated to be $X_{\rm esc}\simeq2.0(E_p/250\rm~GeV)^{-\delta}\rm~g~cm^{-2}$, where $\delta=0.46$ for $E_p<250$ GeV and $\delta=0.33$ for $E_p>250$ GeV \citep{PAMELA14a,AMS16a,2019PhRvD..99f3012M}. Then, the CR escape rate from the ISM is estimated to be $E_pU_{E_p}V_{\rm gal}/t_{\rm conf}\approx{E_p}U_{E_p}cM_{\rm gas}/X_{\rm esc}$, where $U_{E_p}$ is the differential energy density of CR protons and $M_{\rm gas}\simeq8\times10^{9}M_\odot$ is the total gas mass in our Galaxy \citep{NS16a}. This escape rate should balance with the injection rate, and then, we can estimate the CR proton intensity, $\Phi_p=cU_{E_p}/(4\pi{E_p})$, to be \citep{KMM18a}
\begin{equation}
 E_p^2\Phi_p\approx\frac{E_pQ_{E_p}X_{\rm esc}}{4\pi{M_{\rm gas}}}.
\end{equation}

Figure \ref{fig:CR} depicts the CR proton spectrum from the QBXB-MADs. Our scenario can reproduce the CR proton data around the knee energy within the uncertainty ranges of the experimental data and the total number of QBXB-MADs. The CR composition around the knee is dominated by protons \citep{2020arXiv201210372T}, which is also consistent with our prediction. Galactic SNRs should account for CRs of $\lesssim10^6$ GeV, while other sources, such as binary neutron-star merger remnants \citep{KMM18a} or past activities of Sgr A* \citep{FMK17a}, should be responsible for CRs of $\gtrsim3\times10^7$ GeV.

\section{Discussion}\label{sec:discussion}

  \begin{figure}
   \begin{center}
    \includegraphics[width=\linewidth]{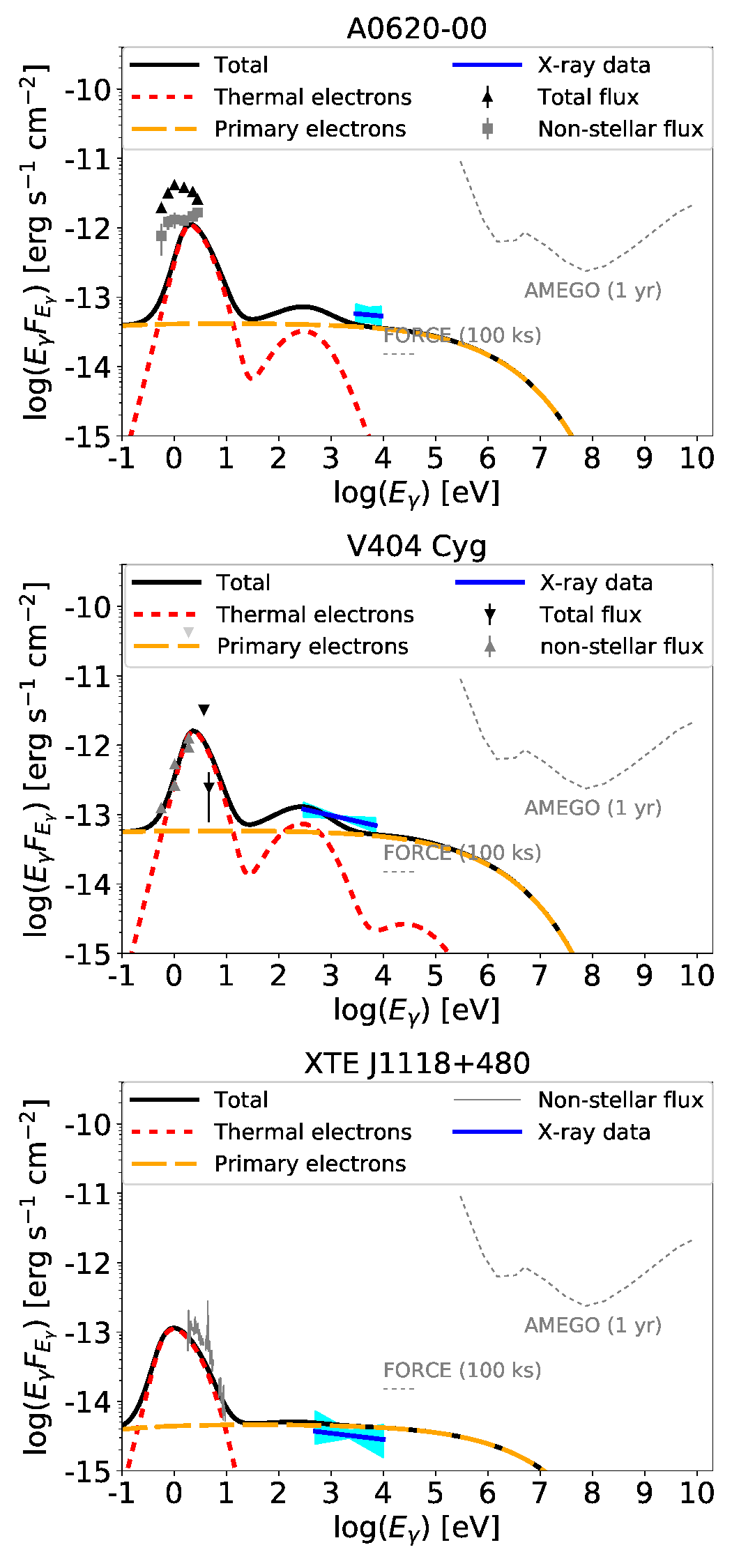}
    \caption{Same as Figure \ref{fig:spe}, but for $s_{\rm inj}=2.0$.}
    \label{fig:s2spe}
   \end{center}
  \end{figure}

  \begin{figure}
   \begin{center}
    \includegraphics[width=\linewidth]{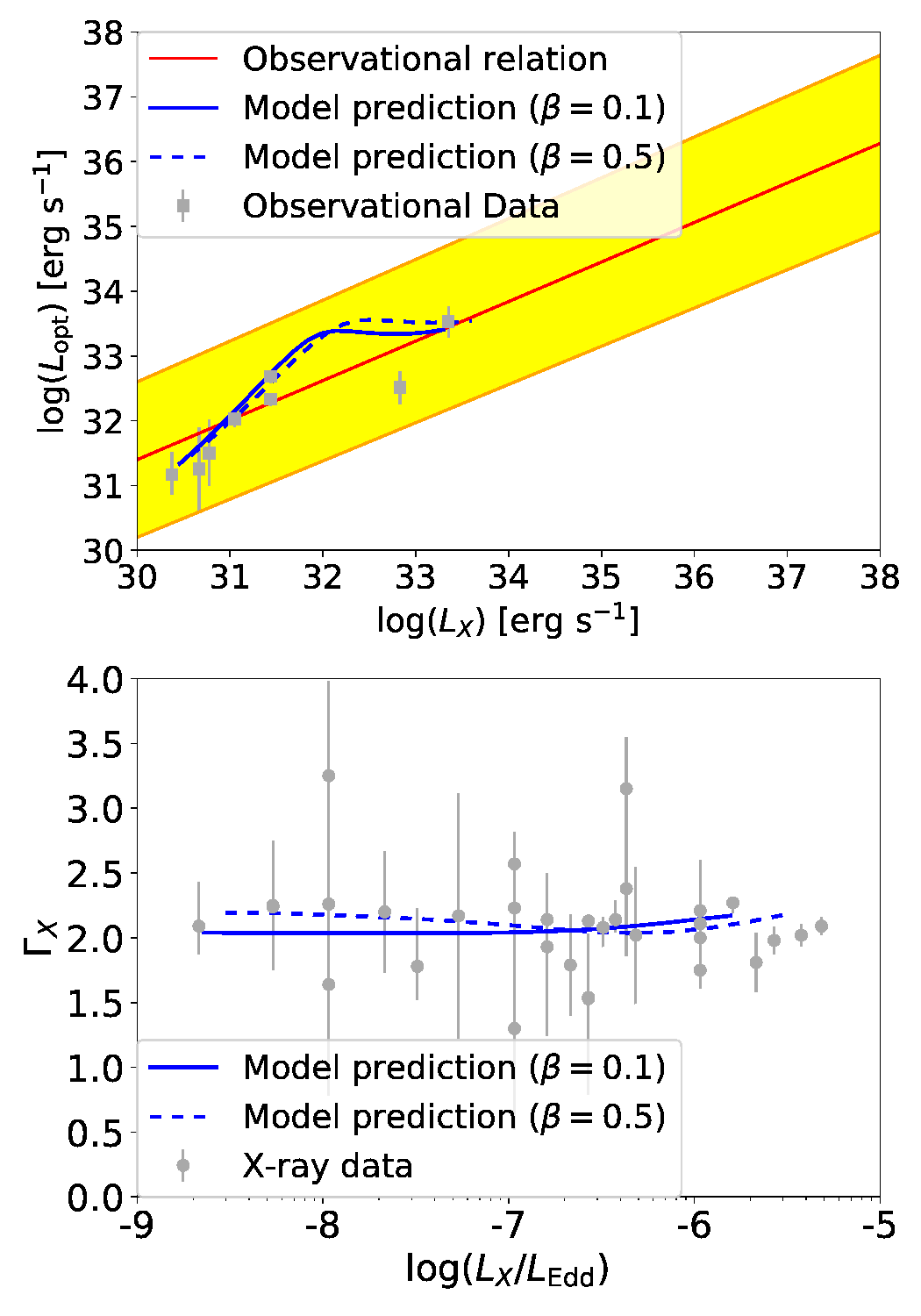}
    \caption{Same as Figure \ref{fig:Loir}, but for $s_{\rm inj}=2.0$.}
    \label{fig:s2Loir}
   \end{center}
  \end{figure}

  \begin{figure}
   \begin{center}
    \includegraphics[width=\linewidth]{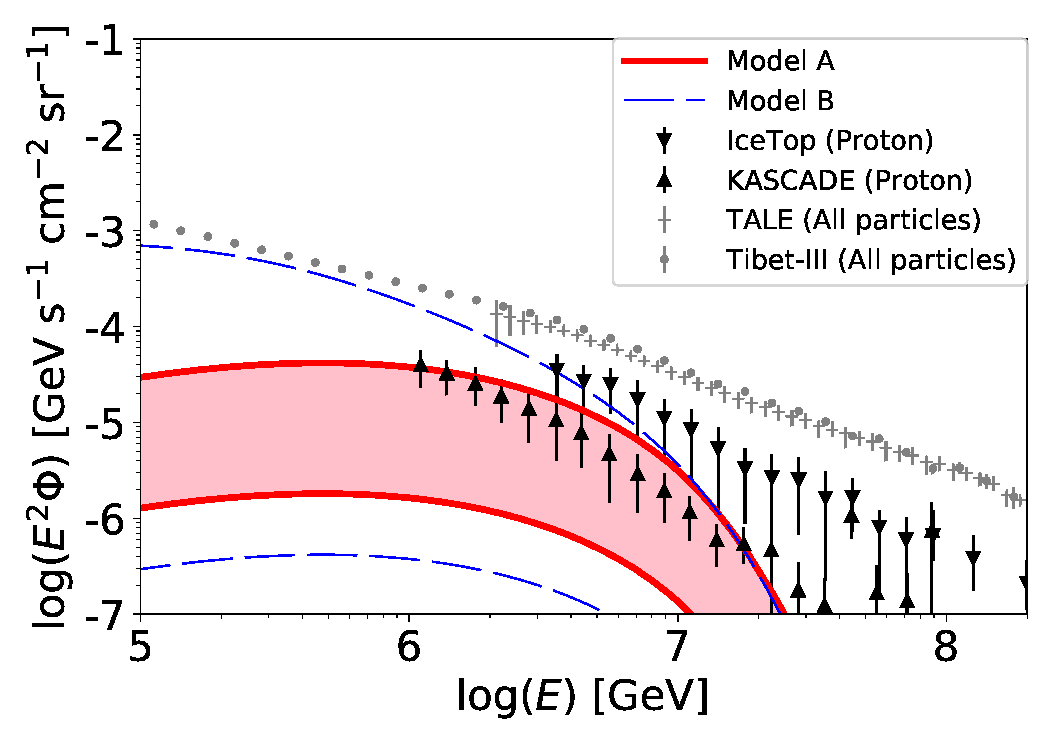}
    \caption{Same as Figure \ref{fig:CR}, but for $s_{\rm inj}=2.0$.}
    \label{fig:s2CR}
   \end{center}
  \end{figure}

\subsection{Differentiating the emission models for QBXBs}

Our QBXB-MAD scenario is distinguishable from the previously proposed scenarios. Moderately magnetized RIAFs, or standard and normal evolution (SANE) scenarios, usually produce bumpy spectra \citep{EMN97a,1997ApJ...482..448N}. We calculate the spectra by the SANE-mode RIAFs with one-zone approximation in Figure \ref{fig:spe}, whose parameters are $(\alpha,~\dot{m}/10^{-4})=$ (0.2, 0.70), (0.2, 1.0), (0.1, 0.50) for A0620-00, V404 Cyg, and XTE J1118+480, respectively. We set $\beta=3.0$ for all the objects. The SANE-mode RIAFs lead to very soft spectra in the hard X-ray range. On the other hand, MADs produce power-law hard X-ray spectra owing to their non-thermal electrons. FORCE will easily discriminate this feature with $30-100$ ks integration for A0620-00 and V404 Cyg. XTE J1118+480 demands more time, and the predicted flux by the MAD scenario is close to the design sensitivity. The jet scenarios \citep{ycn05,2008MNRAS.389..423P} also produce a power-law X-ray spectrum. The time lag of the variability in different wavelengths can be useful to differentiate the scenarios. We expect that the time lag between X-ray and optical bands are shorter than that in the X-ray and radio bands in our QBXB-MAD scenario. This is because the X-ray and optical emissions originate from the same region, while the radio signals originate from jets, and thus, they should follow the variability of X-ray and optical signals with some delay. On the other hand, optical signals are followed by X-ray and radio signals in the jet scenario in which optical emission originates from outer accretion disks. 

\subsection{Cases with higher $\dot{m}$}

We focus on the cases with $\dot{m}_o\lesssim0.01$ in Section \ref{sec:MAD}. For a higher accretion rate of $\dot{m}_o\gtrsim0.1$, the truncation radius given by Equation (\ref{eq:Rtrn}) shrinks to $R_{\rm trn}\lesssim(m_p/m_e)R_G$. This leads electron temperature to the trans-relativistic regime, where our assumption is no longer valid. In this situation, the electron temperature can be almost independent of radius, and we can approximate $t_{pe}\propto N_p^{-1}\propto R^{3/2-s_w}$. For the case without outflows, i.e., $s_w=0$, the dependence of $t_{pe}$ on $R$ is the same with that of $t_{\rm fall}$, and hence, there is no truncation radius at which $t_{pe}$ and $t_{\rm fall}$ balance each other~\citep{mq97}. With the effect of outflows, the truncation radius for a higher $\dot{m}_o$ is smaller than that for a lower $\dot{m}_o$. This feature is consistent with the expectation of X-ray transient observations, where the truncation radius is smaller when the X-ray luminosity is higher \citep{EMN97a}.

For a smaller truncation radius of $R_{\rm trn}\lesssim100R_G$, the magnetic field amplification by advection is not so drastic, and MADs are not formed instantaneously. Even in this case, the poloidal magnetic field can be accumulated at the horizon as the BH keeps  accreting plasmas, and a MAD would eventually be formed if the polarity of the advected poloidal field is aligned for a long time. However, MRI changes the polarity of the field within a few tens of rotation time at the truncation radius \citep[e.g.][]{SI09a}. Then, the accumulated flux may be canceled out before reaching the MAD state if the truncation radius is small. Such a situation is expected in low-hard states. We leave the quantitative estimate of this phenomenon as a future work, because it has no influence on our discussion as long as we focus on the quiescent state.

\subsection{CR production in accretion flows}

CRs are highly likely produced in hot components of accretion flows, such as RIAFs or magnetized coronae above accretion disks \citep{1977ApJ...218..247L,1991ApJ...380L..51H}. Observations of high-energy particles provide direct hints of CR production in accretion flows. First, the infrared and X-ray flares of Sgr A* is likely produced by non-thermal electrons accelerated in RIAFs \citep[see e.g.,][for reviews]{2010RvMP...82.3121G}. Also, Fermi-LAT detected GeV gamma-rays from a few radio-quiet active galactic nuclei (AGN) \citep{WNX15a,2020ApJ...892..105A}, which may imply the non-thermal particle production in RIAFs or coronae. In addition, the hottest spot of the IceCube point source search of high-energy neutrinos is associated to NGC 1068, a nearby Seyfert galaxy \citep{Aartsen:2019fau}. If this neutrino signal is real, the neutrino flux is much higher than the upper limit of the TeV gamma-rays \citep{2019ApJ...883..135A}, which indicates that the neutrino source should be ``hidden'' in gamma-rays \citep{Murase:2015xka}. This implies that the high-energy neutrinos should be produced at the accretion corona rather than the star-burst activities \citep{2019ApJ...880...40I,Murase:2019vdl}.

From the theoretical view points, the hot components consist of collisionless plasma where the Coulomb relaxation timescale is longer than the viscous dissipation timescales \citep{tk85,mq97,ktt14,kmt15,Murase:2019vdl}. Accretion flows should be turbulent due to MRI in SANEs \citep{bh91,SP01a,MM03a,2012MNRAS.426.3241N} or MRTI in MADs \citep{MTB12a,2019ApJ...874..168W}. Then, CRs are naturally accelerated  by magnetic reconnections \citep{hos15,KSQ16a,2020PhPl...27h0501G} and/or stochastic acceleration by turbulence \citep{2012SSRv..173..535P,KTS16a,2019MNRAS.485..163K,2018ApJ...867L..18Z,2018PhRvL.121y5101C}.
Recent particle-in-cell (PIC) simulations indicated that the particle acceleration is efficient in a strongly magnetized plasma of magnetization parameter $\sigma\approx B^2/(4\pi m_p c^2)>1$ \citep{2014ApJ...783L..21S,2016ApJ...818L...9G}\footnote{Protons in accretion flows are usually non-relativistic, and we can use the cold plasma limit in the expression of $\sigma$.}. Based on GR MHD simulations, MADs can supply reconnection layers of $\sigma\gtrsim1$ \citep{BOP18a,2020ApJ...900..100R}, and hence, MADs likely produce CRs efficiently. \cite{2020ApJ...905..178K} demonstrated that non-thermal particles accelerated in MADs can account for the gamma-ray emissions from nearby radio galaxies. 

In our scenario, we assume a hard spectral index of CR protons and electrons, $s_{\rm inj}=1.3$. Although magnetic reconnections in solar flares seem to produce a softer CR spectrum \citep{2002SSRv..101....1A,2021ApJS..252...13A}, we expect a hard CR spectrum in MADs because magnetic reconnections in highly magnetized plasma $(\sigma>1)$ can efficiently produce non-thermal particles whose spectral index is as hard as $s_{\rm inj}\sim1-2$ according to PIC simulations \citep{2014ApJ...783L..21S,2016ApJ...818L...9G}. Recent results may indicate that the spectrum can be $s_{\rm inj}\simeq2$ if non-thermal particles are accelerated to higher energies of $E_i\gtrsim \sigma m_i c^2$ \citep{2018MNRAS.481.5687P}. Our scenario can explain the optical and X-ray data in quiescent states with $s_{\rm inj}=2$, as demonstrated in Figures \ref{fig:s2spe} and \ref{fig:s2Loir}.  A softer CR spectrum results in a lower CR proton luminosity at PeV energies. Nevertheless, the resulting CR proton intensity with $s_{\rm inj}=2$ is still fairly consistent with the knee-energy CRs as seen in Figure \ref{fig:s2CR}. 

PIC simulations need to resolve gyration scales of thermal particles, which is several orders of magnitude smaller than the scale of turbulence generated by MRTI. We need to follow the turbulence generated at the largest scale, i.e., the MAD scale, in order to understand the particle acceleration of PeV protons. Since simulations that can follow both the accretion flow and gyration scales are impossible with current facilities, the particle spectra realized in MADs are still uncertain. Further theoretical and observational studies are necessary to unravel the particle accelerations at the vicinity of BHs.

\subsection{Relativistic jets from MADs}

MADs can launch relativistic jets, whose power is estimated to be $L_j\approx\epsilon_j\dot{m}L_{\rm Edd}\simeq 1.3\times10^{35}\epsilon_{j,-1}\dot{m}_{-3}M_1$, where $\epsilon_j$ is the jet production efficiency. For a rapidly spinning BH of $a\approx1$, where $a$ is the dimensionless spin parameter, $\epsilon_j\approx1$ can be achieved, while $\epsilon_j\lesssim0.1$ is more appropriate for a moderate spin parameter of $a\lesssim0.5$ \citep{TNM11a,MTB12a,2019ApJ...875L...5E}. The Hillas energy for protons in the jets can be estimated to be \citep{Hil84a,LW09a,IMT17a}
\begin{equation}
 E_{\rm Hil}\approx \frac{2e}{\Gamma_j\theta_j}\sqrt{\frac{L_j}{c}}\simeq 1.1 L_{j,35}^{1/2}(\Gamma_j\theta_j)^{-1} \rm~PeV
\end{equation} 
where $\Gamma_j$ is the jet Lorentz factor and $\theta_j$ is the jet opening angle. Thus, the jets from QBXB-MADs are also potential PeVatrons. The CR luminosity can be estimated to be $L_{\rm CR}^{\rm jet}\approx \epsilon_{\rm CR}L_j$. The ratio of CR luminosities of jets to MADs is estimated to be $L_{\rm CR}^{\rm jet}/L_{\rm CR}^{\rm MAD}\approx (\epsilon_j\epsilon_{\rm CR})/(\epsilon_{\rm dis}\epsilon_{\rm NT})$. For slowly spinning BHs, we expect $\epsilon_j\epsilon_{\rm CR}\lesssim0.01$, which results in $L_{\rm CR}^{\rm jet}/L_{\rm CR}^{\rm MAD}<1$. On the other hand, rapidly spinning BHs can achieve $\epsilon_j\epsilon_{\rm CR}\gtrsim0.1$, and then, the CR production can be dominated by the jets. Therefore, the spin parameter measurement is crucial to clarify the CR production sites in our scenario. 

In hard states, BH X-ray binaries show a correlation between the X-ray luminosity and the radio luminosity, $L_R$ \citep{2003MNRAS.345.1057M,2004A&A...414..895F}. This correlation may hold in quiescent states \citep{2020ApJ...889...58R}. In our scenario, the radio emission is produced by jets, while X-rays originate from MADs. Since both the jet power and the X-ray luminosity from MADs are proportional to the mass accretion rate, it is natural to have the $L_R-L_X$ correlation in QBXB-MADs. The jet production efficiency depends on the spin parameter, and thus, Galactic BHs should have a similar value of $a$ in order to hold a tight correlation. Since the $L_X-L_R$ relation in QBXBs are still not solid, we leave further quantitative discussion as a future work. 

BH spins can be measured by the spectral fitting of disk blackbody radiation and the iron fluorescence line broadened by the gravitational redshift \citep[see][for reviews]{2003PhR...377..389R,2006ARA&A..44...49R,2007A&ARv..15....1D}. The X-ray data from Cyg X-1 and GX 339-4 can be interpreted with a very high spin of $a\gtrsim0.9$ \citep{2008ApJ...679L.113M,2011ApJ...742...85G}, whereas other interpretations with lower spin values are possible \citep[e.g.][]{2009ApJ...707L.109Y,2010MNRAS.406.2206K,2017PASJ...69...36K}. The high resolution X-ray spectroscopy by XRISM \citep{2020arXiv200304962X} will be able to measure the spin parameters more accurately. These observations should be done during outbursts, because the X-ray fluxes in quiescent states are too low for XRISM to perform spectroscopic observations.
On the other hand, gravitational-wave observations revealed that typical binary BHs have a low value of $a$ \citep{Abbott:2020niy}. Although the progenitor of binary BHs should be high-mass X-ray binaries that is a distinct population from low-mass X-ray binaries, this may indicate that stellar mass BHs typically have a low value of $a$ without significant dispersion. In this case, CR proton production in MADs would dominate over that in jets, and QBXB-MADs would likely reproduce the $L_R-L_X$ correlation.

\section{Summary}\label{sec:summary}

We have discussed the formation scenario of MADs in QBXBs, demonstrated that their broadband photon spectra are consistent with those observed in selected QBXBs, and proposed them as Galactic PeVatrons. In a quiescent state, the mass accretion rate onto a BH is so low that the accretion flow cannot cool efficiently, leading to formation of a RIAF at $\sim10^4R_G$. Then, strong outflows produce poloidal fields, and a fast infall will carry the magnetic flux to the  vicinity of the BH. 
We conclude that the accreting plasma with weak poloidal fields ($\beta_p\sim10^4$) at $R_{\rm trn}\sim10^4R_G$ can form a MAD. Future MHD simulations covering the region from the vicinity of the BH to the truncation radius will be able to test this scenario.

Since the MAD consists of a strongly magnetized collisionless plasma, CR electrons  are naturally accelerated via magnetic reconnections or stochastic acceleration by turbulence. The magnetic reconnections and turbulence cascades also heat up the plasma, and hence, the plasma in the MAD consists of both thermal and non-thermal components. Thermal electrons emit infrared and optical photons via the cyclosynchrotron emission, and CR electrons produce broadband photons from X-rays to MeV gamma-rays by the synchrotron process. We demonstrated that QBXB-MADs can explain both the infrared/optical and X-ray data of the selected BH binaries that have rich data in the quiescent states. QBXB-MADs can also reproduce the observational correlations: $L_X-L_{\rm opt}$ and $L_X/L_{\rm Edd}-\Gamma_X$.

CR protons are also accelerated in the QBXB-MADs, and those with TeV-PeV energies can diffusively escape from the system. We have estimated the number density of the QBXBs by two methods, and calculated the CR energy density in the ISM using the grammage and total gas mass in the Galaxy. We found that the CR protons escaping from the QBXB-MADs are fairly consistent with the CR intensity of PeV energies.

In the QBXB-MAD scenario, non-thermal electrons produce a hard photon spectrum in the hard X-ray to MeV gamma-ray ranges. Future hard X-ray satellites, such as FORCE, will provide a good test of our model. The variability time lags in the different wavelengths (radio/optical/X-rays) can be useful to distinguish our QBXB-MAD model from jet models. Also, if we find an appropriate target with a high accretion rate of $\dot{m}\gtrsim10^{-3}$, future MeV satellites will be able to detect the QBXB-MADs. These observations are crucial to understand the nature of the QBXBs, especially non-thermal phenomena at the vicinity of BHs. These data, together with the multi-TeV gamma-ray data by future detectors, such as LHASSO \citep{2019arXiv190502773B}, CTA \citep{2019scta.book.....C}, and SWGO \citep{2019arXiv190208429A}, will enable us to identify cosmic PeVatrons. 


\acknowledgements
This work is partly supported by JSPS Research Fellowship and KAKENHI No. 19J00198 (S.S.K.), 18J20943 (T.S.), 20K04010, 20H01904 (K.K), and Hakubi project at Kyoto University (N.K.).

\bibliographystyle{aasjournal}
\bibliography{ssk}

\listofchanges

\end{document}